\documentclass{PoS}
\usepackage{epsfig}
\usepackage{amsmath}
\usepackage{amssymb}
\usepackage{amsbsy}
\usepackage{amsfonts}
\usepackage{graphics}
\usepackage{latexsym}
\usepackage{mathrsfs}
\usepackage{dcolumn}
\usepackage{wasysym}

\vspace*{-2cm}

\title{\hfill {\large \tt DESY 08-159}\\[-0.5em]\hfill {\large \tt Edinburgh
    2008/44}\\[-0.5em]\hfill {\large \tt LTH 810}\\[2em] 
Extracting the $\rho$ resonance from lattice QCD simulations at small
  quark masses}

\ShortTitle{Extracting the $\rho$ resonance}

\author{M. G\"ockeler$^a$, R. Horsley$^b$, Y. Nakamura$^c$, D. Pleiter$^c$,
  P.E.L. Rakow$^d$, \speaker{G. Schierholz}$^{\,c a}$ and J. Zanotti$^b$\\ \\
        $^a$ Institut f\"ur Theoretische Physik, Universit\"at Regensburg,
        93040 Regensburg, Germany\\
        $^b$ School of Physics and Astronomy, University of Edinburgh,
        Edinburgh EH9 3JZ, UK\\
        $^c$ Deutsches Elektronen-Synchrotron DESY, John von Neumann Institut
        f\"ur Computing NIC, 15738 Zeuthen, Germany\\
        $^d$ Theoretical Physics Division, Department of Mathematical
        Sciences, University of Liverpool, Liverpool L69 3BX, UK}
\author{QCDSF Collaboration}

\abstract{Using established relations between the scattering matrix in
          infinite volume and the two-particle spectrum in a periodic box,
          we compute the mass and width of the $\rho$ meson from simulations of
          $N_f=2$ flavors of dynamical clover fermions at small pion masses $2
          m_\pi < m_\rho$.}  

\FullConference{The XXVI International Symposium on Lattice Field Theory\\
                 July 14-19 2008\\
                 Williamsburg, Virginia, USA}

\begin{document}

\section{Introduction}

Lattice simulations of QCD with dynamical fermions have reached the point now
where the masses of up and down quarks are light enough to allow the low-lying
resonances, such as the $\rho$ and $\Delta$, to decay via the strong
interactions. The problem of extracting masses and widths of unstable 
particles from the lattice data is complicated by the fact that resonance
states cannot be identified with a single energy level of the lattice
Hamiltonian. For elastic two-body resonances the method of choice, originally
proposed by L\"uscher~\cite{Luscher} and Wiese~\cite{Wiese}, is to compute the
phase shift in the infinite volume from the volume dependence of the energy
spectrum, and from that the mass and width of the resonance. In this talk we
shall present first results of an attempt to compute mass and width of the
$\rho$ meson from simulations at realistic quark masses. 

The $\rho$ meson is practically a two-pion resonance. It has isospin 1,
and the two pions form a $p$-wave state. We denote the pion momentum in the
center-of-mass frame by $k=|\vec{k}|$. Phenomenologically, the scattering phase
shift $\delta_{11}(k)$ is very well described by the effective range formula
\begin{equation}
\frac{k^3}{W}\, \cot\,\delta_{11}(k) = \frac{24\pi}{g_{\rho\pi\pi}^2}\,
\left(k_\rho^2 - k^2\right)\;,
\label{effective}
\end{equation}
where $\displaystyle W=2\sqrt{k^2+m_\pi^2}$ and $\displaystyle k_\rho=
\frac{1}{2}\sqrt{m_\rho^2-4m_\pi^2}$\,. The width of the $\rho$ is given by  
\begin{equation}
\Gamma_\rho=\frac{g_{\rho\pi\pi}^2}{6\pi} \, \frac{k_\rho^3}{m_\rho^2}\;.
\end{equation}
Experimentally $\Gamma_\rho = 146 \, \mbox{MeV}$, which translates into
$g_{\rho\pi\pi}=5.9$. The phase shift $\delta_{11}(k)$ passes  through $\pi/2$
at the physical $\rho$ mass, {\it i.e.} $k^2=k_\rho^2$ or
$W^2=m_\rho^2$.  

In the case of noninteracting pions, the possible energy levels in a periodic
box of length $L$ are given by 
\begin{equation}
W=2\sqrt{k^2+m_\pi^2}\;,
\label{energy}
\end{equation}
where $k=2\pi |\vec{n}|/L$, $\vec{n}$ being a vector with components $n_i \in
\mathbb{N}$. In the interacting case, the energy levels are still given by
(\ref{energy}), but now $k$ is the solution of~\cite{Luscher} 
\begin{equation}
\delta_{11}(k) = \arctan\left\{\frac{\pi^{3/2}
    q}{\mathcal{Z}_{00}(1,q^2)}\right\}\: \mbox{mod}\; \pi\;, \quad
    q=\frac{kL}{2\pi}\;, 
\label{phase}
\end{equation}
where $\mathcal{Z}_{00}$ is a generalized zeta
function~\cite{Luscher2}. That is to say,   
each energy value $W$, computed on the periodic lattice at some fixed value of
$m_\pi$, gives rise to a certain momentum $k$. The scattering phase at this
momentum and pion mass is given by (\ref{phase}). Fitting $\delta_{11}(k)$
to the effective range formula (\ref{effective}) then allows us to estimate
the mass and width of the $\rho$ meson.

In Fig.~1 we show the expected energy spectrum at the physical pion mass. The
lowest energy level lies significantly below the physical $m_\rho$ 
(indicated by the dashed line), and should not be mistaken for the $\rho$ 
mass. In Fig.~2 we show the same plot, but for
$m_\pi/m_\rho=0.3$, corresponding to a pion mass of $m_\pi \approx 300 \,
\mbox{MeV}$. In this case the lowest energy level differs only little from
$m_\rho$.

\begin{figure}[t]
\vspace*{-2.0cm}
\begin{center}
\epsfig{file=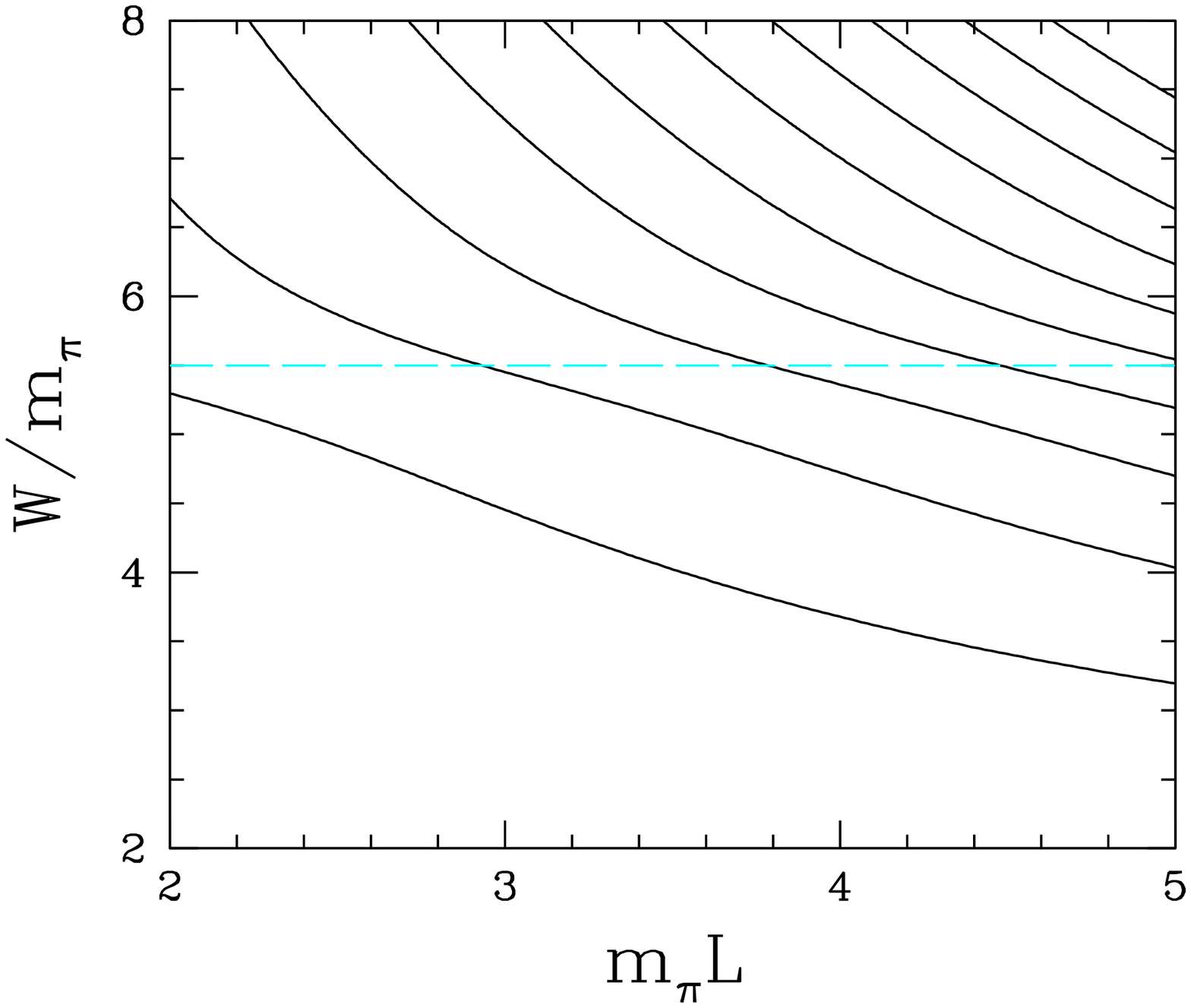,width=10cm,clip=}
\caption{The expected energy spectrum in the $\rho$ channel for physical
  values of $m_\pi$ and $m_\rho$. The dashed line shows the ratio
  $W/m_\pi=m_\rho/m_\pi=5.5$.}  
\end{center}
%
\vspace*{-1cm}
\begin{center}
\epsfig{file=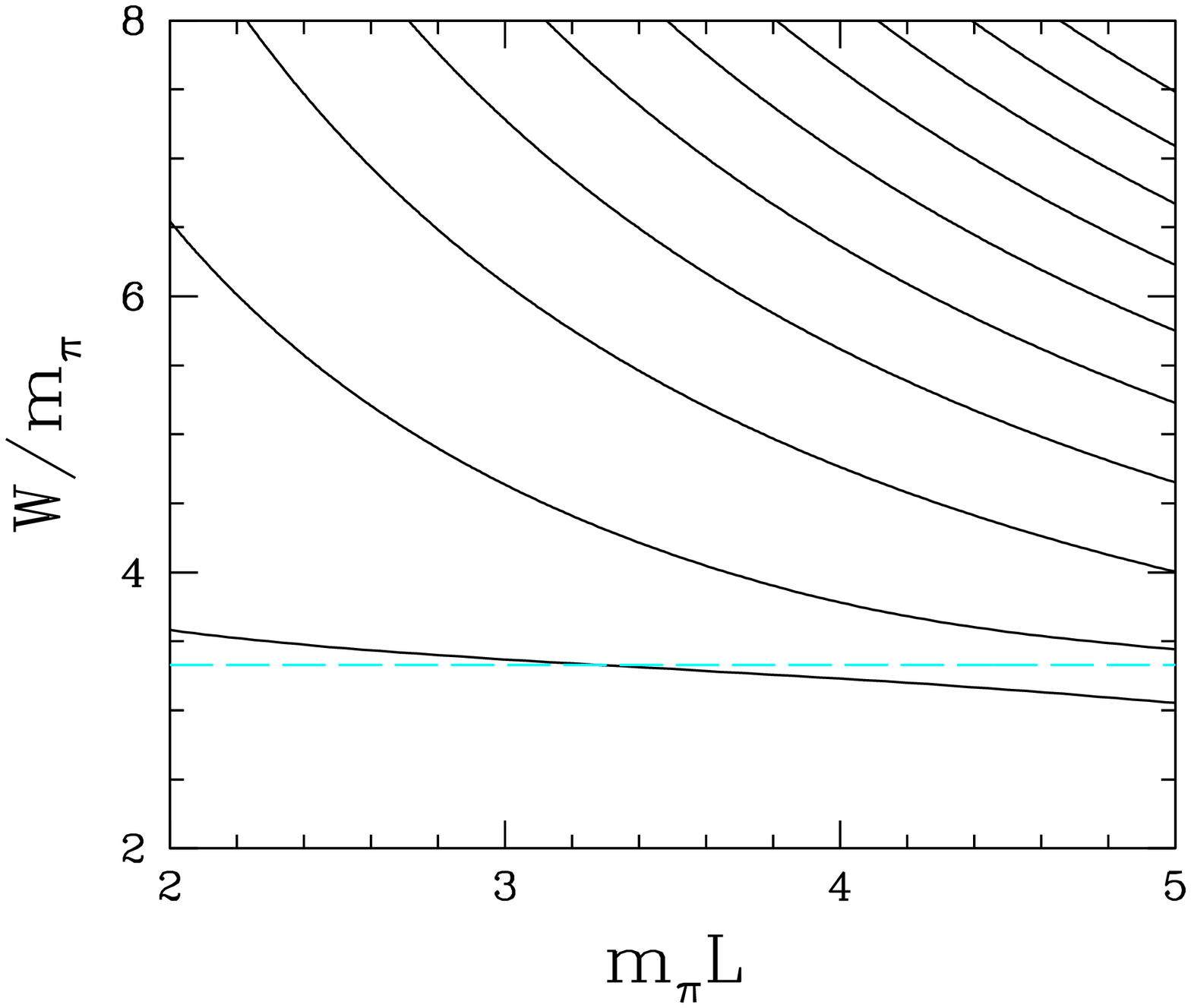,width=10cm,clip=}
\caption{The expected energy spectrum in the $\rho$ channel for $m_\rho/m_\pi =
  3.33$. The dashed line shows the ratio $W/m_\pi=m_\rho/m_\pi=3.33$.}    
\end{center}
\end{figure}

\clearpage
\begin{figure}[t]
\vspace*{-2.0cm}
\begin{center}
\epsfig{file=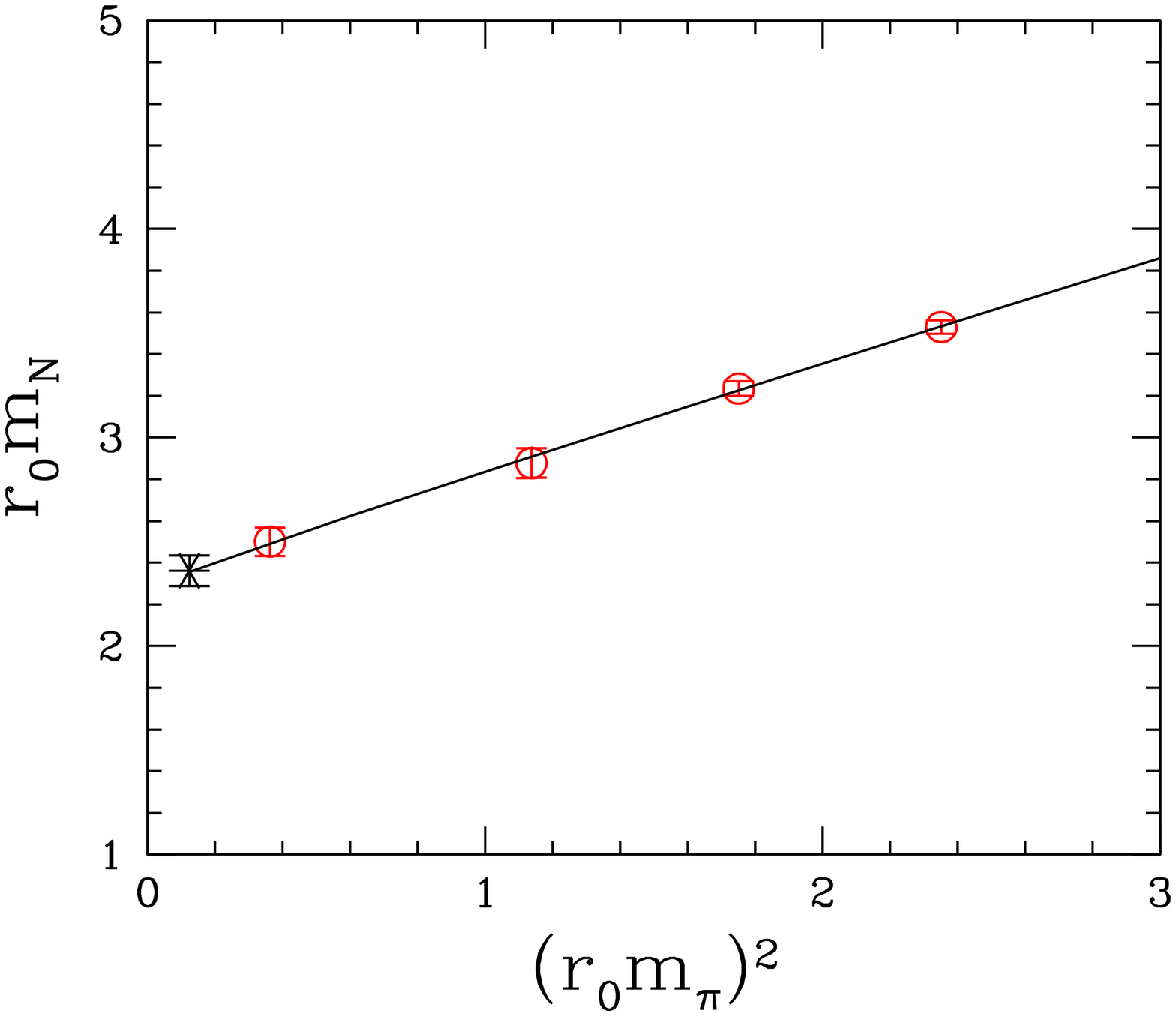,width=10cm,clip=}
\caption{The nucleon mass as a function of the pion mass squared at
  $\beta=5.40$, together with the chiral extrapolation.}  
\end{center}
%
\vspace*{-1cm}
\begin{center}
\epsfig{file=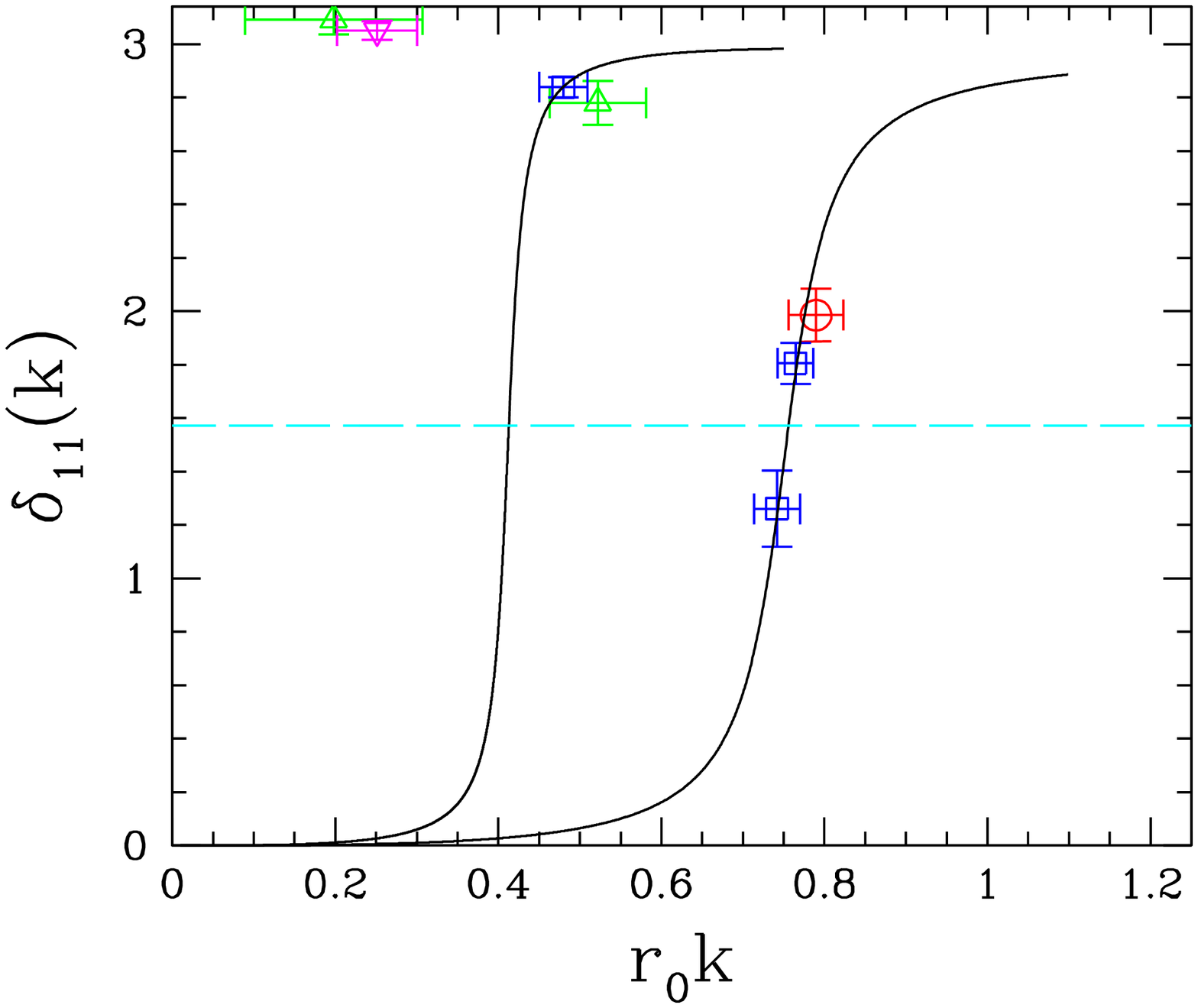,width=10cm,clip=}
\caption{The phase shift $\delta_{11}(k)$ as a function of $r_0 k$, together
  with the effective range fit from Fig.~5. The symbols indicate the different
  couplings: $\beta=5.25$ ($\triangledown$), $\beta=5.29$ ($\square$),
  $\beta=5.30$ ($\vartriangle$), and $\beta=5.40$ ($\Circle$). The
  $\beta=5.30$ data points have been inferred from~\cite{DelDebbio}. The
  curves refer to pion masses of $m_\pi=250$ (right) and $390$ MeV (left),
  respectively.}     
\end{center}
\end{figure}

\clearpage
\section{Lattice simulation and results}

The simulations reported here use nonperturbatively $O(a)$ improved Wilson
fermions with $N_f=2$ degenerate flavors of dynamical quarks. The parameters
of our current data sample are listed in Table~1. In five cases the pions are
light enough so that the $\rho$ can decay. These data sets are marked by a 
$\star$. We use the nucleon mass to
set the scale. In Fig.~3 we show the chiral extrapolation of $m_N r_0$ to the
physical pion mass for our finest lattice at $\beta=5.40$. A
fit~\cite{AliKhan} to the physical nucleon mass, which now includes pion 
masses below $300\,\mbox{MeV}$, gives $r_0=0.495(19)\, \mbox{fm}$. We shall
use this number throughout this paper. We do not see any scaling violations
outside the error bars.  

We restrict the analysis to the lowest energy level. To compute $W$, we use
Jacobi-smeared sources and sinks, as well as wall sources and smeared sinks,
and employ the lattice dispersion relation
\begin{equation}
\left(2\sinh \frac{W}{4}\right)^2 = k^2+m_\pi^2
\end{equation}
to determine $k$. The phase shift is then readily obtained from
(\ref{phase}). In Fig.~4 we show $\delta_{11}(k)$ as a function of $r_0 k$. We
also include two data points from~\cite{DelDebbio} ($D_3$ and $D_4$) on the
$24^3\times 48$ lattice at $\beta=5.30$. These authors use the same action as
ours. The scale parameter $r_0/a$ at this $\beta$ value has been obtained by
interpolation. The phase shift clearly depends on the mass of the pion. In the
following we shall combine the three data points at our lowest pion masses,
$m_\pi = 240 - 250 \, \mbox{MeV}$, to a single data set and treat the
remaining data points separately. 

\begin{table}[t]
\begin{center}
\begin{tabular}{c|c|c|c|c}
$\beta$ & $\kappa_{\rm sea}$ & Volume & $a$ [fm] & $m_\pi$ [MeV] \\
\hline
 5.25    & 0.13575 & $24^3\times 48$ & 0.084 & 600 \; \phantom{$\star$}\\
 5.25    & 0.13600 & $24^3\times 48$ & 0.084 & 430 \; $\star$\\
\hline
 5.29    & 0.13550 & $24^3\times 48$ & 0.080 & 810 \; \phantom{$\star$}\\
 5.29    & 0.13590 & $24^3\times 48$ & 0.080 & 590 \; \phantom{$\star$}\\
 5.29    & 0.13620 & $24^3\times 48$ & 0.080 & 390 \; $\star$\\
 5.29    & 0.13632 & $32^3\times 64$ & 0.080 & 250 \; $\star$\\
 5.29    & 0.13632 & $40^3\times 64$ & 0.080 & 250 \; $\star$\\
\hline
 5.40    & 0.13500 & $24^3\times 48$ & 0.072 & 810 \; \phantom{$\star$}\\
 5.40    & 0.13560 & $24^3\times 48$ & 0.072 & 770 \; \phantom{$\star$}\\
 5.40    & 0.13610 & $24^3\times 48$ & 0.072 & 610 \; \phantom{$\star$}\\
 5.40    & 0.13625 & $24^3\times 48$ & 0.072 & 530 \; \phantom{$\star$}\\
 5.40    & 0.13640 & $24^3\times 48$ & 0.072 & 420 \; \phantom{$\star$}\\
 5.40    & 0.13640 & $32^3\times 64$ & 0.072 & 420 \; \phantom{$\star$}\\
 5.40    & 0.13660 & $32^3\times 64$ & 0.072 & 240 \; $\star$\\
\end{tabular}
\caption{Volumes, lattice spacings and pion masses of the data sample used in
  the analysis. A $\star$ indicates that $m_\rho > 2 m_\pi$\,.} 
\end{center}
\end{table}

To obtain $m_\rho$ and $\Gamma_\rho$, we fit the data to the effective range
formula (\ref{effective}). This is done in two steps. We first fit the 
three data points in the lowest mass bin. This gives $m_\rho$ and
$g_{\rho\pi\pi}$ at the corresponding pion mass. Assuming that $g_{\rho\pi\pi}$
does not depend on the pion mass,
\begin{figure}[h!]
\vspace*{-1.5cm}
\begin{center}
\epsfig{file=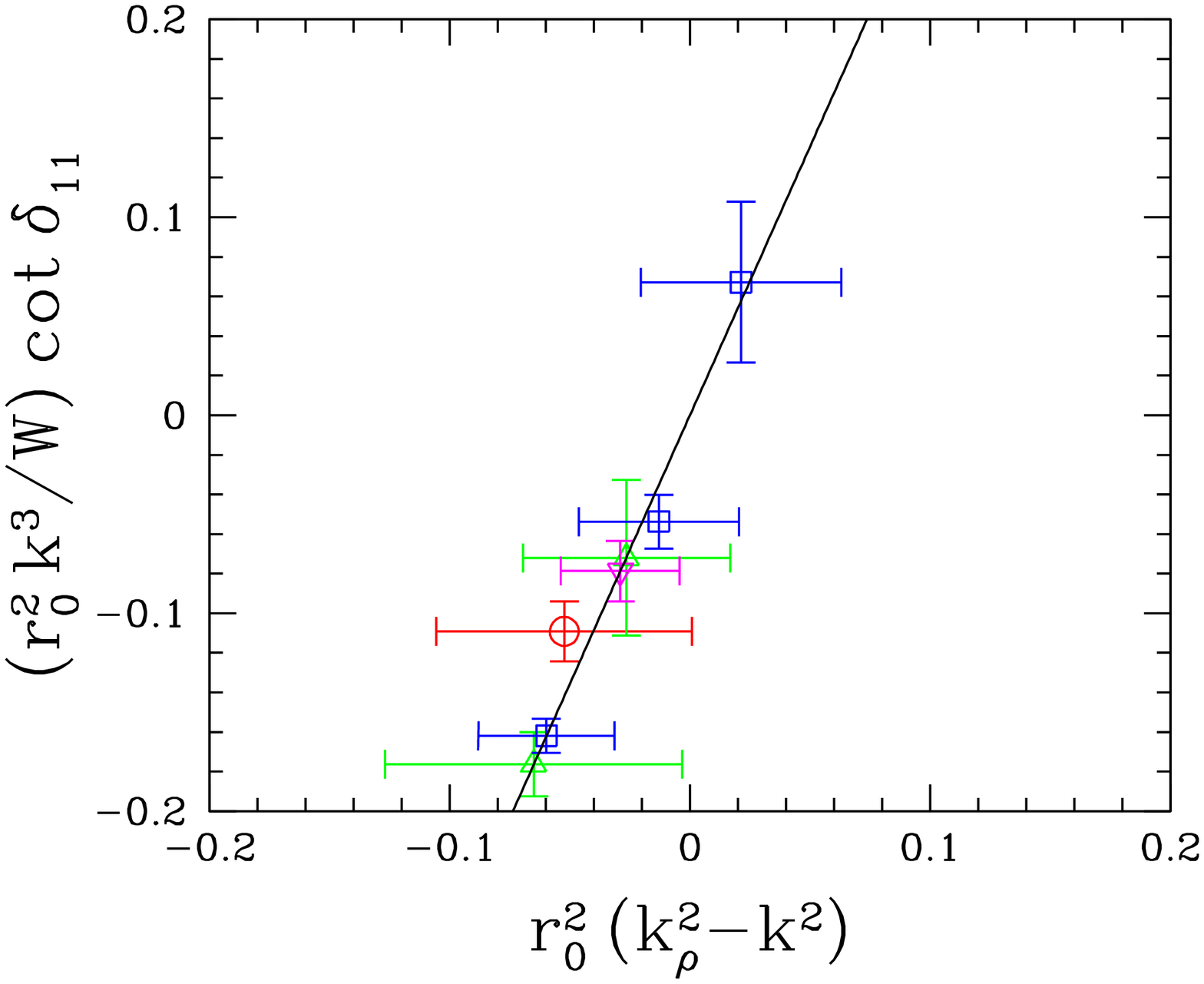,width=10cm,clip=}
\caption{The fit of $\delta_{11}(k)$ from Fig.~4 to the effective range
  formula (1.1). The top two 
  squares and the circle refer to the data points in the lowest mass bin. The
  symbols are as in Fig.~4.}   
\end{center}
%
\vspace*{-1cm}
\begin{center}
\epsfig{file=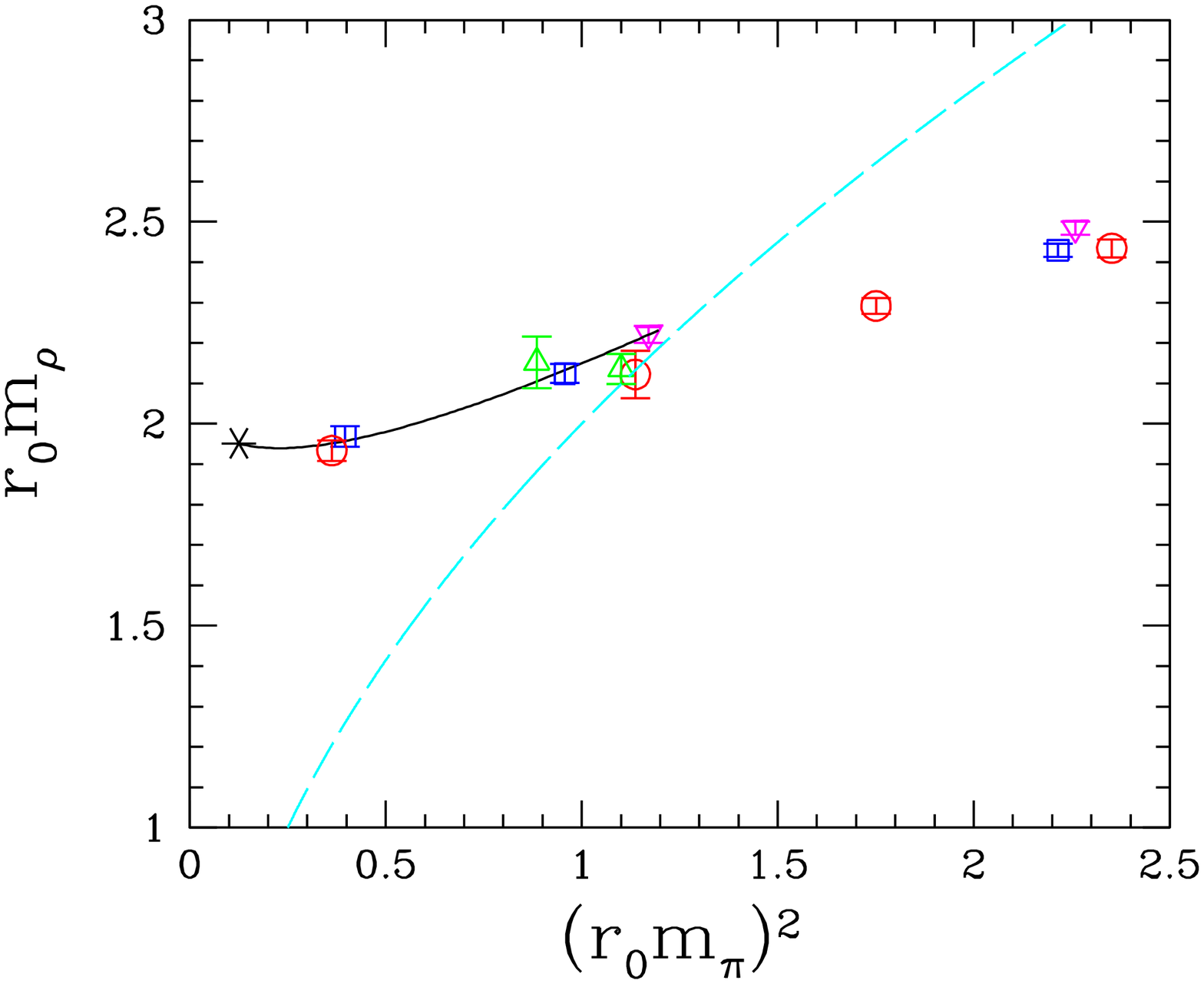,width=10cm,clip=}
\caption{The true $\rho$ mass as a function of the pion mass squared. The
  solid curve is a chiral fit~\cite{Bruns} through the physical $\rho$
  mass ($\ast$). The dashed line indicates the two-pion threshold. The symbols
  are as in Fig.~4.}    
\end{center}
\end{figure}
we subsequently fit the remaining data
points. In Fig.~5 we show the result of the fit, and in Fig.~4 we comparethe resulting phase shift $\delta_{11}(k)$ with the original data. For the
$\rho \rightarrow \pi\pi$ 
coupling we obtain 
\begin{equation}
g_{\rho\pi\pi} = 5.3_{\,-1.5}^{\,+2.1} \;,
\end{equation}
which is in broad agreement with the phenomenological value. The $\rho$ masses
obtained from the fit are plotted in Fig.~6, together with the rest of the
masses lying above the two-pion threshold. The solid curve is a fit through
the experimental $\rho$ mass. It appears that $m_\rho$ cannot be analytically
continued across the two-pion threshold.

\section{Conclusions and outlook}

So far we have concentrated on the lowest energy level only. To compute the
energies of states with higher relative momenta, we need to extend the basis
of operators to operators which especially project onto two-pion scattering
states~\cite{Gockeler,Aoki}. With this information it should be possible to
trace out the phase shift over the entire elastic region~\cite{Gockeler} $0 <
k < \sqrt{3} \, m_\pi$\,. At present we are performing simulations at lighter
pion masses. Altogether, this should enable us to present a 
more precise calculation of $\displaystyle g_{\rho\pi\pi}$, including an
extrapolation to the chiral limit.

In the conference presentation an attempt was made to compute the mass and
width~\cite{Bernard} of the $\Delta$ resonance. It was found that this
requires pion masses close to the physical value, as the $\pi N$ threshold is
only reached at $(r_0 m_\pi)^2 \approx 0.5$, as opposed to $(r_0 m_\pi)^2
\approx 1$ in case of $\rho \rightarrow \pi\pi$ (see Fig.~6). We hope to
return to that in due course.  

\section*{Acknowledgment}

The numerical simulations have been performed on the Altix at LRZ (Munich),
the BlueGeneLs at EPCC (Edinbugh) and NIC (J\"ulich), as well as on the
APEmille and apeNEXT at NIC (Zeuthen). This work is 
supported by the EU Integrated Infrastructure Initiative Hadron Physics (I3HP)
under contract RII3-CT-2004-506078 and by the DFG under contracts FOR 465
(Forschergruppe Gitter-Hadronen-Ph\"anomenologie) and SFB/TR 55 (Hadron
Physics from Lattice QCD).


\begin{thebibliography}{99}

\bibitem{Luscher}
  M.~L\"uscher,
  Commun.\ Math.\ Phys.\  {\bf 105} (1986) 153;
  Nucl.\ Phys.\  B {\bf 364}, 237 (1991).

\bibitem{Wiese}
  U.-J.~Wiese,
  Nucl.\ Phys.\ Proc.\ Suppl.\  {\bf 9}, 609 (1989).

\bibitem{Luscher2}
  M.~L\"uscher,
  Nucl.\ Phys.\  B {\bf 354}, 531 (1991).

\bibitem{AliKhan}
  A.~Ali Khan {\it et al.},
  Nucl.\ Phys.\  B {\bf 689}, 175 (2004)
  [arXiv:hep-lat/0312030].

\bibitem{DelDebbio}
  L.~Del Debbio {\it et al.},
  JHEP {\bf 0702}, 082 (2007)
  [arXiv:hep-lat/0701009].

\bibitem{Bruns}
  P.~C.~Bruns and U.~G.~Meissner,
  Eur.\ Phys.\ J.\  C {\bf 40}, 97 (2005)
  [arXiv:hep-ph/0411223].

\bibitem{Gockeler}
  M.~G\"ockeler {\it et al.},
  Nucl.\ Phys.\  B {\bf 425}, 413 (1994)
  [arXiv:hep-lat/9402011].

\bibitem{Aoki}
  S.~Aoki {\it et al.},
  Phys.\ Rev.\  D {\bf 76}, 094506 (2007)
  [arXiv:0708.3705 [hep-lat]].

\bibitem{Bernard}
  V.~Bernard {\it et al.},
  Eur.\ Phys.\ J.\  A {\bf 35}, 281 (2008).

\end{thebibliography}
\end{document}